\documentclass[traditabstract]{aa} 

\usepackage{graphicx}
\usepackage{txfonts}
\begin{document}
   \title{The BaSTI Stellar Evolution Database: models for extremely metal-poor and super-metal-rich stellar populations}

\authorrunning{Pietrinferni et al.}
\titlerunning{Extension of the BaSTI model archive}

\author{Adriano Pietrinferni  \inst{1}, Santi Cassisi \inst{1}, Maurizio Salaris \inst{2}, \and Sebastian Hidalgo \inst{3}}

\institute{INAF - Osservatorio Astronomico di Teramo, Via M. Maggini, 64100 Teramo, Italy;\\
 \email{cassisi@oa-teramo.inaf.it, pietrinferni@oa-teramo.inaf.it}
 \and
Astrophysics Research Institute, Liverpool John Moores University,Liverpool Science Park, 146 Brownlow Hill, 
IC2 Building, Liverpool L3 5RF \\
 \email{M.Salaris@ljmu.ac.uk}
\and
Instituto Astrofisico de Canarias, C.lle Via Lactea, sn., La Laguna (Tenerife), Spain,\\
\email{shidalgo@iac.es}
} 

   \date{Received  ; accepted }
   
\abstract{
We present an extension of the BaSTI stellar evolution database to extremely metal poor (${\rm Z=10^{-5}}$) and 
super-metal-rich (Z=0.05) metallicities, with both scaled-solar and $\alpha$-enhanced ([$\alpha$/Fe]=0.4) 
heavy element distributions. These new tracks (from the pre-main sequence to the early-asymptotic giant branch phase), horizontal branch models   
and isochrones, will enable the use of the BaSTI database to study, i.e., the most metal poor populations found 
in Local Group faint dwarf galaxies, and the metal rich component of the Galactic bulge.
An overview of several fundamental predictions of stellar evolution over the full metallicity range of BaSTI is presented, 
together with comparisons with literature calculations at ${\rm Z=10^{-5}}$ and Z=0.05.
}

\keywords{galaxies: stellar content -- stars: evolution -- stars: interiors  -- stars: horizontal-branch} 
\maketitle

\section{Introduction}

\noindent
Libraries of stellar models and isochrones covering wide ranges of age and metallicities are an essential tool to investigate the
properties of resolved and unresolved stellar populations. This is the seventh paper of a series devoted to create an extended, 
complete, and up-to-date database of theoretical models and byproducts such as isochrones, 
luminosity functions (LF), tables of integrated spectra and magnitudes 
of simple stellar populations, and synthetic Colour-Magnitude-Diagrams (CMDs) 
of simple/composite stellar systems (BaSTI -- a Bag of Stellar 
Tracks and Isochrones\footnote{Available at the URL site http://www.oa-teramo.inaf.it/BASTI}). 
The first paper (Pietrinferni et al.~2004 -- Paper~I) presented scaled-solar 
stellar evolution models and isochrones, while Paper~II (Pietrinferni et al.~2006) extended the database to 
$\alpha$-enhanced metal compositions, appropriate, for example, to model the stellar population of the Galactic halo 
(Gratton, Sneden \& Carretta~2004, and references therein).  
The inclusion of the asymptotic giant branch phase, and then description of the BaSTI 
synthetic CMD generator (SYNTHETIC\_MAN) was included in 
Cordier et al.~(2007 -- Paper III), whilst integrated spectra and magnitudes for the whole set of chemical compositions 
were published by Percival et al. (2009 -- Paper IV). 
Pietrinferni et al. (2009 -- Paper V) extended BaSTI to mixtures including the CNONa abundance anticorrelations 
observed in Galactic globular clusters (GCs), while 
Salaris et al.~(2010 -- Paper VI) added to the database white dwarf WD cooling sequences and isochrones. 
As part of our ongoing effort to provide the scientific community with a self-consistent evolutionary framework to interpret 
observations of a wide range of stellar populations, we now present an extension of the 
BaSTI archive to both extremely metal-poor and super-metal-rich stellar populations, beyond the present 
BaSTI metallicity range.

Observations of metal-poor candidates in the HK survey (see, e.g., Christlieb et al.~2008 and references therein), 
the Sloan Digital Sky Survey (York et al. 2000) and the SEGUE survey (Yanny et al.~2009) 
have vastly increased the available sample of Galactic extremely metal-poor stars, with ${\rm [Fe/H]<-3.5}$ (see also Norris et al.~2013). 
Needless to say, the interpretation of the chemical and evolutionary properties of these 
stars can provide vital clues on the earliest phases of the formation and evolution of the Galaxy. 

At the same time, detailed investigations of the smallest galaxies in the Local Group, the so-called
Ultra Faint Dwarfs (Belokurov et al.~2007, 2010, Norris et al.~2010) have shown that these stellar systems host a sizable, if not dominant, 
population of very metal-poor objects. Observations have disclosed the presence of stars 
with [Fe/H] as low as $\approx -$3.7 (see, e.g., Fulbright et al.~2004, Norris et al.~2008,~2010).
To infer the star formation history of these Milky Way satellites, population synthesis codes 
have to account for the observed low metallicities.

Extended sets of stellar models for extremely metal-poor stars have been calculated by Cassisi \& Castellani~(1993), 
Cassisi, Castellani \& Tornamb\'e~(1996) and Cassisi, Castellani \& Castellani~(1997); these calculations 
have enabled detailed investigations of the properties of 
extremely metal-poor objects and primordial
stellar populations (Raimondo, Brocato \& Cassisi~2001). 
However, the significant improvements of the last decade regarding stellar physics inputs 
(see Cassisi 2005, 2009, 2010, 2012 and references therein) require 
an updated theoretical framework for these populations.

\begin{figure*}
 \centering
 \includegraphics[scale=0.7]{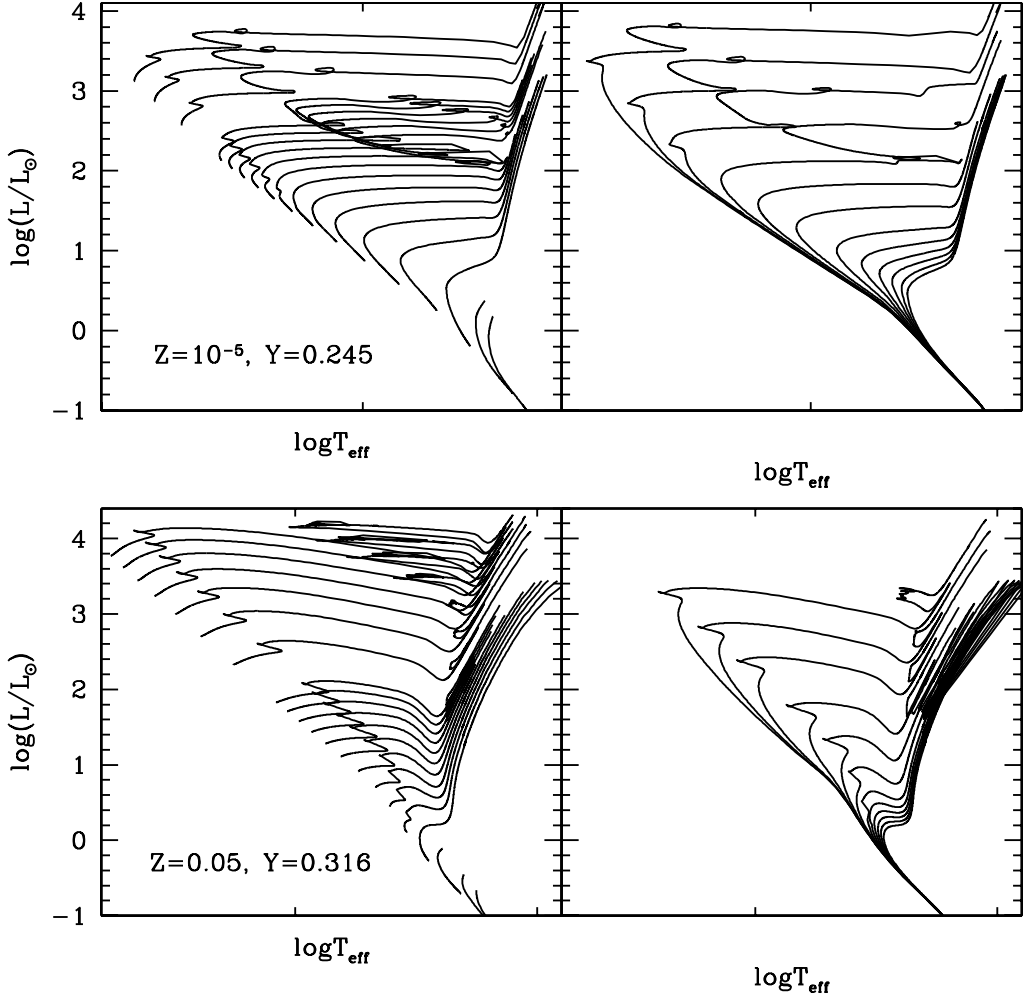}
 \caption{\textit{Upper left panel}: evolutionary tracks of selected stellar models 
(no convective core overshooting during the H-burning phase) with ${\rm Z=10^{-5}}$, Y=0.245, 
${\rm M/M_\odot: 0.5, 0.6, 0.8, 1.0, 1.2, 1.4, 1.6, 1.8, 2.0, 2.2, 2.4, 2.6, 2.8, 3.0, 4.0, 5.0, 6.0}$, and 
a scaled solar heavy elements mixture.
All tracks for low-mass stars (apart from those masses with lifetime much longer than the Hubble time) 
are displayed up to the tip of the red giant branch (RGB). \textit{Upper right panel}:
 selected isochrones for the same chemical composition, with ages 
${\rm t(Gyr)=0.05, 0.10, 0.20, 0.5, 1.0, 2.0, 4.0, 6.0, 8.0, 10.0, 12.0, 14.0}$.
\textit{Lower left panel}: as the upper left panel but for Z=0.05, Y=0.316. Masses equal to 
${\rm M/M_\odot: 7, 8, 9, 10}$ are also shown. 
\textit{Lower right panel}: as the upper right panel but for the metal-rich chemical composition.}
\label{fig:trkiso}
\end{figure*}

At the other extreme of the metallicity spectrum, 
both spectroscopic and photometric surveys (see Gonzalez et al.~2011 and references therein) show that  
the Galactic bulge metallicity distribution (peaked at about solar metallicity) displays an extended tail 
reaching ${\rm [Fe/H]\sim+1.0}$. High-metallicity stars are also an important constituent of elliptical galaxies, 
and their post-Main Sequence (post-MS) evolution is considered to be responsible (Greggio \& Renzini~1990) for the observed UV excess --  
the so called UV-upturn phenomenon (Burstein et al.~1988 and references therein) -- observed in the spectra of a fraction of elliptical galaxies. 
From a theoretical point of view, during the last 15 years calculations of 
very metal-rich models have received increasing attention (see, i.e., Bono et al. 1997, Dotter et al.~2007), and we 
therefore extend BaSTI to cover also models at the upper end of the observed cosmic metallicities.

\begin{figure*}
 \centering
 \includegraphics[scale=0.63]{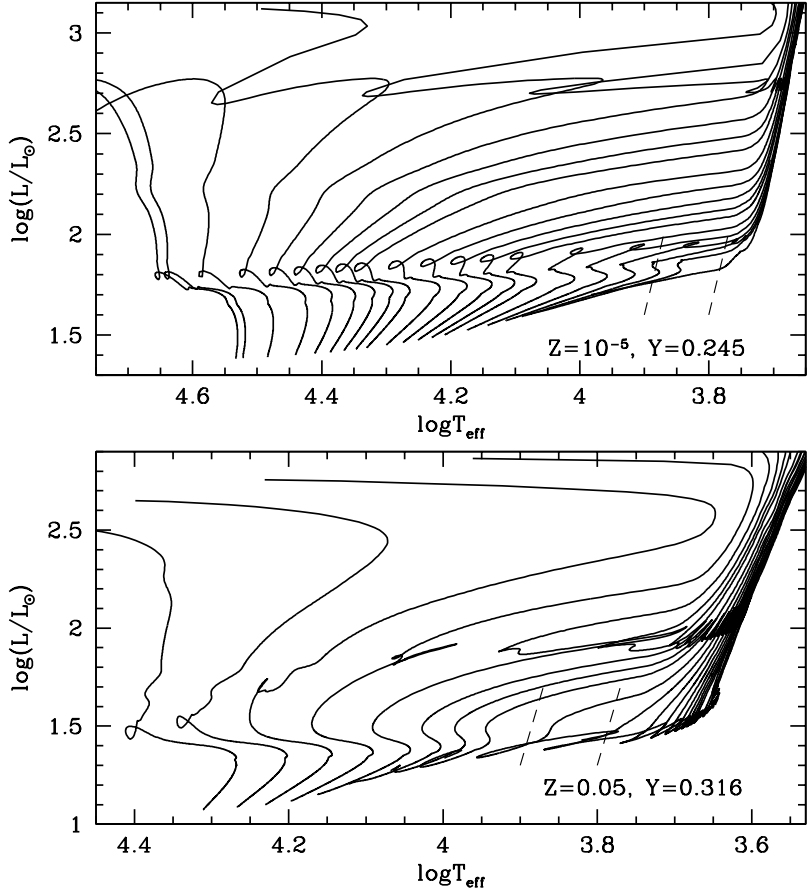}
 \caption{\textit{Upper panel}: evolutionary tracks for selected HB 
models originated from a ${\rm 0.8M_\odot}$ RGB progenitor with scaled solar ${\rm Z=10^{-5}}$, Y=0.245. 
From left to right the tracks are for masses ${\rm M/M_\odot: 0.517, 0.518, 0.520}$, then in steps of 
${\rm 0.005M_\odot}$ up to ${\rm 0.55M_\odot}$, in steps of ${\rm 0.01M_\odot}$ up to ${\rm 0.6M_\odot}$, 
in steps of ${\rm 0.02M_\odot}$ up to ${\rm 0.7M_\odot}$, and then ${\rm \Delta{M}=0.1M_\odot}$ up to ${\rm 0.8M_\odot}$. 
The approximate positions of the blue and red boundaries of the RR Lyrae instability strip are  
also displayed as thin, dashed lines.
 \textit{Lower panel}: as the upper panel but for Z=0.05, Y=0.316 and a RGB progenitor with mass equal to ${\rm 1M_\odot}$. 
The less massive HB model has a mass equal to ${\rm 0.465M_\odot}$. 
The mass steps are as in the upper panel, but for masses larger than ${\rm 0.6M_\odot}$, where ${\rm \Delta{M}=0.1M_\odot}$.}
\label{fig:hb}
\end{figure*}

The extension of BaSTI to both extremely metal-poor and super-metal-rich compositions will complete the creation of 
an updated and self-consistent
theoretical evolutionary framework that covers (almost) the whole metallicity interval spanned by stars in both the  
Galaxy and extra-galactic stellar systems.
 
The paper is organized as follows: \S~2 summarizes the model physics inputs and describes 
calculations and main evolutionary properties. Comparisons with models 
available in literature are discussed in \S~3. A summary and final remarks follow in \S~4.

\section{The theoretical framework}
 
To be consistent with the existing BaSTI calculations, we used the same stellar evolution code (see Paper I, II and III), 
and input physics of the previous calculations (see Paper I, II and III) e.g., 
radiative and electron conduction opacities, equation of state, and nuclear cross sections. 
Superadiabatic convection is treated according to the Cox \& Giuli~(1968) formalism of the mixing 
length theory  (B\"ohm-Vitense~1958), with the  
the mixing length value equal to 2.023, 
as obtained from a calibration of the standard solar model (SSM)\footnote{The first release of the BaSTI database relied 
on the low-temperature opacities by Alexander \& Ferguson~(1994) that required a mixing length value equal to 1.913 for the SSM. 
When the new low-T opacities by Ferguson et al. (2005) were released, they were adopted to 
recompute the whole BaSTI archive; the SSM required in this case a mixing length equal to
2.023.}  Outer boundary conditions have been computed by integrating the 
atmospheric layers with the ${\rm T(\tau)}$ relation by Krishna-Swamy~(1966).

All models include mass loss using the Reimers formula (Reimers 1975) with the free parameter 
$\eta$=0.4 \footnote{The BaSTI archive provides 
calculations for two different values fof the free paramter $\eta$, namely 0.2 and 0.4. 
Our preferred value for population synthesis 
models is $\eta=0.4$. We also noticed that this value has become a \lq{standard}\rq\ 
choice among BaSTI users, and therefore here 
we decided to neglect the $\eta=0.2$ computations.}. 
For consistency with the other computations already available at the BaSTI 
database, all stellar models
presented in this work have been computed by neglecting the occurrence of atomic diffusion.

As for the other calculations already available in the BaSTI library, we computed models with and without 
convective core overshoot during the H-burning phase (when convective cores are present).  
The extension of the overshooting region as a function of mass is exactly the same adopted in our previous calculations.

The extremely metal-poor chemical compositions presented here have Y=0.245 (Cassisi, Salaris \& Irwin~2003), 
total metallicity ${\rm Z=10^{-5}}$ 
and both a scaled solar metal mixture (as in Paper~I we rely on the Grevesse \& Noels~1993 mixture) 
and an $\alpha$-enhanced one (see Paper~II for more details). The scaled solar 
calculations correspond to ${\rm [Fe/H]=-3.27}$, whilst the $\alpha$-enhanced ones to ${\rm [Fe/H]=-3.62}$.

The super-metal-rich calculations are for Y=0.316 (corresponding to ${\rm \Delta{Y}/\Delta{Z}\sim1.4}$, as in the 
existing BaSTI calculations) and Z=0.05, both scaled solar and $\alpha$-enhanced. 
These choices correspond to ${\rm [Fe/H]=+0.51}$ (scaled solar) and $+0.16$ ($\alpha$-enhanced).
The adopted ${\rm \Delta{Y}/\Delta{Z}\sim1.4}$ for the supersolar regime is 
corroborated by the good match of BaSTI isochrones to the CMD of the old open cluster NGC6791 
([Fe/H]$\sim$=0.4) and a derived distance modulus (${\rm (m-M)_v=13.50}$ -- Bedin et al. 2008) 
in perfect agreement with the eclipsing binary distance to the cluster determined by Brogaard et al. (2012). 

For each chemical composition we have computed stellar models with mass ranging from $0.50M_\odot$ 
to $10M_\odot$\footnote{For ${\rm Z=10^{-5}}$ the most massive stellar model corresponds to $6M_\odot$.}. 
All models -- with the exception of those with a MS lifetime longer than the Hubble time for which the computation has ben stopped at the central H exhaustion -- 
have been computed from the pre MS until the beginning of the thermal pulse stage along the asymptotic giant branch.
These have been supplemented by an extended set of horizontal branch (HB) calculations, by adopting the He-core mass 
and envelope He abundance of a red giant branch (RGB) progenitors whose age at He-ignition is $\approx$13~Gyr 
($0.8M_\odot$ for ${\rm Z=10^{-5}}$ and $1.0M_\odot$ for Z=0.05). 
These models allow to compute synthetic HB populations with 
an arbitrary mass distribution.
Figure~\ref{fig:trkiso} displays the theoretical Hertzsprung-Russell (HR) diagram of selected tracks and isochrones for both metallicities, 
whilst Fig.~\ref{fig:hb} shows HB tracks.

The evolutionary tracks have been reduced to the same number of points -- to facilitate the computation of isochrones 
and their use in population synthesis codes -- by identifying along each evolutionary track some characteristic 
homologous points (key points -- KPs) 
corresponding to well-defined evolutionary phases.  
For a careful description of the adopted KPs we refer the reader to Paper~II. The whole set of evolutionary 
computations (but the additional HB
models discussed above) have been used to compute isochrones from 30~Myr (50~Myr for 
${\rm Z=10^{-5}}$) to 15~Gyr. 
Finally, tracks and isochrones have been transformed to
various photometric systems (i.e., Johnson-Cousins, ACS and WFC3 Vegamag)  
by using the same colour-${\rm T_{eff}}$ transformations and bolometric corrections presented 
in Paper~I and II, and Bedin et al. (2005). 
These computations are made public at the BaSTI official website http://www.oa-teramo.inaf.it/BASTI. 

\begin{figure}
 \centering
 \includegraphics[scale=0.45]{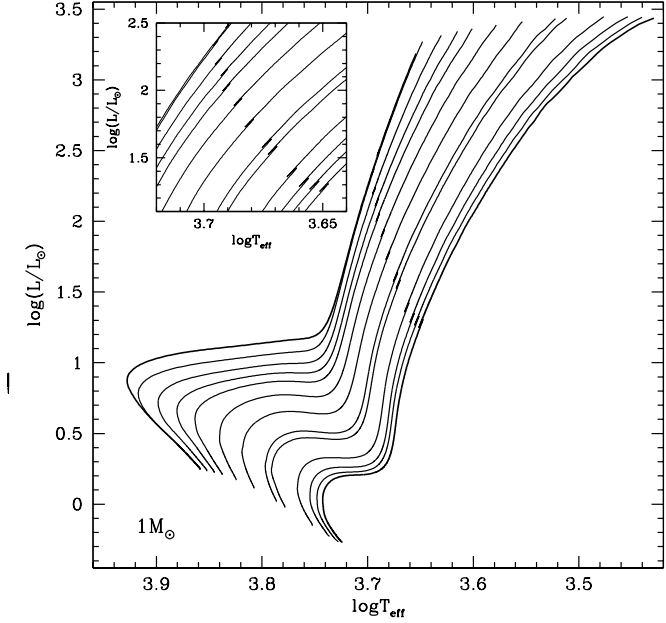}
 \caption{HR diagrams of ${\rm 1M_\odot}$ models, 
for the full set of BaSTI scaled solar metallicities: (from left to right) Z=$10^{-5}$, $10^{-4}$,
 $3\times10^{-4}$, $6\times10^{-4}$, 0.001, 0.002, 0.004, 0.008, 0.01, 0.0198, 0.03, 0.04 and 0.05. 
The inset shows a zoom on the portion of the RGB bump region.}
\label{fig:trkzz}
\end{figure}

\subsection{Main evolutionary properties of extremely metal-poor and super-metal-rich stars}

Figure~\ref{fig:trkiso} reveals some important evolutionary features of these models, related 
to their extreme metallicities. 
At fixed initial mass, extremely metal-poor stars are obviously much hotter and brighter than 
their more metal-rich counterparts. This occurrence is, as well known, due to the huge dependence of
the stellar radiative opacity on the metallicity; the larger the metallicity, the larger 
the radiative opacity. 
For an overview across the whole metallicity range spanned by the BaSTI models, 
Tables~\ref{tab:1mevo}  and ~\ref{tab:4mevo}  list 
some relevant evolutionary features for selected models, with Z ranging from ${\rm Z=10^{-5}}$ to 0.05.
 
The strong dependence of the ${\rm T_{eff}}$ scale of low-mass models on Z, associated to the metallicity  
dependence of the radiative opacity, is shown by Fig.~\ref{fig:trkzz}, that 
compares tracks for the same mass, and all (scaled solar) metallicities available in the BaSTI database. 
However, when lowering the metallicity down to values ${\rm Z\le10^{-4}}$, 
the dependence of the radiative opacity on the metal content vanishes, as
demonstrated by the two tracks for ${\rm Z=10^{-4}}$ and ${\rm Z=10^{-5}}$, that overlap almost perfectly along the MS and the 
RGB. This agrees with the results by Cassisi \& Castellani~(1993, and reference therein). 

Something similar occurs for the super-metal-rich models. Although the radiative opacity 
is still affected by the metallicity increase, the effect on both bolometric luminosity 
and effective temperature of an increase ${\rm \Delta{Z}=0.01}$ is larger between, i.e., Z=0.02 and 0.03 
than between Z=0.04 and 0.05.
This is due to the fact that the opacity increase (at fixed T and $\rho$) 
for a fixed ${\rm \Delta Z}$ is larger around ${\rm
Z\sim{Z_\odot}}$ than around Z=0.04 -- because of a larger percentage increase ${\rm \Delta Z/Z}$ --  
as we have verified on the relevant opacity tables. 
One has also to bear in mind an additional effect. 
For the assumed helium-enrichment ratio, the increase of initial He associated to the increase of Z 
has the effect of increasing the model ${\rm T_{eff}}$, compared to an increase of Z at constant Y
(see, e.g. Cassisi \& Salaris~2013). This compensates -- at least partially -- 
the ${\rm T_{eff}}$ decrease associated with the Z increase. 

From data in Table~\ref{tab:1mevo} one notices that, at fixed total mass, the central H-burning lifetime is strongly affected 
by the metal content, monotonically increasing with increasing metallicity as a consequence of the lower
brightness of the stellar structures.

The huge impact of the metallicity on the opacity stratification of the model envelopes  
is also demonstrated by the $\sim$2.5~mag decrease of the RGB bump magnitude, 
when increasing Z from $10^{-5}$ to Z=0.05, as shown in the inset of 
Fig.~\ref{fig:trkzz}. 

It is worth mentioning the trend of the RGB tip luminosity with metallicity.  
The luminosity increases steadily from ${\rm Z=10^{-5}}$ to ${\rm Z\approx0.03}$, followed by a slight decrease 
when the initial metal content increases further. 
On the other hand, the mass of the He core (${\rm M_{cHe}}$) at the He flash decreases monotonically  with increasing 
Z (see Table~\ref{tab:1mevo}). 

\begin{figure}
 \centering
 \includegraphics[scale=0.46]{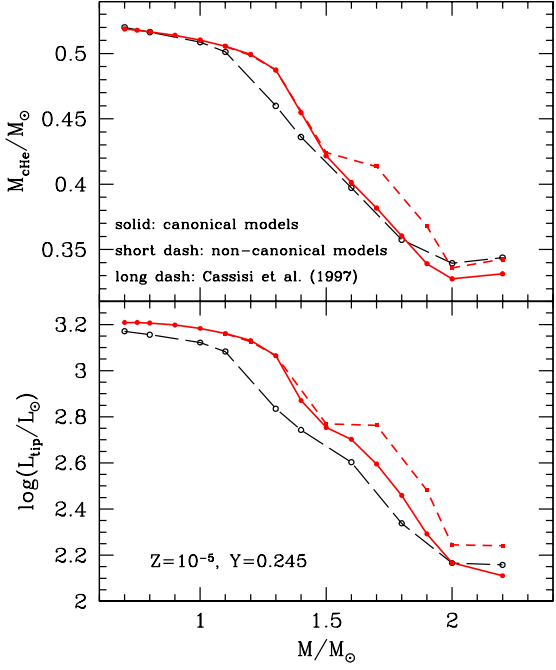}
 \caption{\textit{Upper panel}: behaviour of ${\rm M_{cHe}}$ at the RGB tip as a function of the initial mass 
for the scaled-solar, ${\rm Z=10^{-5}}$ chemical composition, and for both 
canonical (no convective  core overshooting along the MS) and non-canonical (convective core 
overshooting along the MS) stellar models. 
Solid  and short dashed lines correspond to the present calculations, 
the long dashed line to the results by Cassisi et al.~(1997). \textit{Lower panel}: as the upper panel but for the trend of
 the stellar surface luminosity with mass at the RGB tip.}
\label{fig:mchez15}
\end{figure}

\begin{figure}
 \centering
 \includegraphics[scale=0.46]{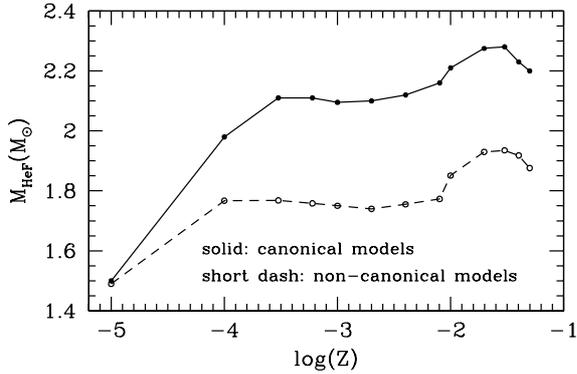}
 \vskip -3cm
 \caption{Behaviour of ${\rm M_{HeF}}$ with metallicity as predicted by BaSTI models for the scaled-solar metal mixture, both with (non canonical models) and without (canonical models) MS convective core overshooting.}
\label{fig:mhefzz}
\end{figure}

One important prediction of stellar model calculations is the transition mass ${\rm M_{HeF}}$ between stars  
that ignite He in an electron degenerate core, and stars that enter the central He burning phase 
without experiencing core electron degeneracy (see, e.g., Sweigart, Greggio \& Renzini 1989, 1990; Cassisi \& Castellani~1993). 
Core mass and bolometric luminosity at He ignition  
change remarkably over a range of only a few tenths of solar mass,  
as shown by Fig.~\ref{fig:mchez15} for the case ${\rm Z=10^{-5}}$, for models 
with and without MS convective core overshooting. 
For this reason, when discussing the evolution of stars with masses around ${\rm M_{HeF}}$, 
one often speaks of a \lq{Red Giant Branch transition}\rq.

The value of ${\rm M_{HeF}}$ depends on the initial chemical composition, as shown by Fig.~\ref{fig:mhefzz}. 
In general, an increase of Z at constant Y would cause a monotonic increase of ${\rm M_{HeF}}$, as a consequence of the 
lower MS luminosity and smaller convective cores (at fixed total mass) during the MS. 
However, the increase of Y with Z due to the adopted ${\rm \Delta{Y}/\Delta{Z}\sim1.4}$ ratio,    
favours the thermal conditions required to ignite He burning, hence a decrease of ${\rm M_{HeF}}$. 
This helps explaining the almost constant value of this parameter in the metallicity range from ${\rm Z\approx3\times10^{-4}}$ 
to $\sim10^{-2}$. At higher Z, up to ${\rm Z\sim0.03}$ the effect associated to the metallicity increase dominates, whilst for 
${\rm Z>0.03}$, the significant He increase forces ${\rm M_{HeF}}$ to decrease.

It is well known that the value of ${\rm M_{HeF}}$ -- for a given chemical composition -- 
depends on the assumed efficiency of the core convective overshooting during the central H-burning stage: 
the larger the overshooting region, the smaller ${\rm M_{HeF}}$ -- see Fig.~\ref{fig:mhefzz}. 
This is due to the larger He core mass at the end of the central 
H-burning stage of the models with MS core overshooting; a larger He core mass at the 
start of the RGB stage implies a hotter core thermal stratification, that favours 
the thermal conditions required for He-burning ignition. 
It is worth noticing how the effect of MS convective core overshooting is vanishing at the 
lowest metallicity. This is due to the huge decrease of the size of the 
convective core during the central H-burning stage in very metal-poor stars, caused by 
the very low efficiency of the CNO-cycle compared to the p-p chain. 

\begin{table*}
\caption{Selected evolutionary properties of a ${\rm 1M_\odot}$ stellar model for all BaSTI metallicities 
(scaled solar mixture). 
${\rm t_H}$ and ${\rm t_{Tip}}$ denote the age of the model at central H exhaustion and at the tip of the RGB, respectively. 
${\rm \Delta{Y_{surf}}}$ is the amount of extra helium dredged to the surface during the ${\rm 1^{st}}$ dredge-up. 
${\rm M_{ZAHB}^{3.85}}$ 
and ${\rm \log(L_{ZAHB}^{3.85}/L_\odot)}$ are the total mass and surface luminosity of the BaSTI HB models whose ZAHB locations 
are at ${\rm \log{T_{eff}}=3.85}$ -- taken as representative of the average effective temperature of 
the RR Lyrae instability strip.}             
\label{tab:1mevo}      
\centering          
\begin{tabular}{c c c c c c c c c c}     
\hline\hline       
                      
Z  & Y &${\rm t_H~(Gyr)}$ & ${\rm \log(L_{Bump}/L_\odot)}$   & ${\rm \log(L_{tip}/L_\odot)}$  & ${\rm M_{cHe}(M_\odot)}$ & ${\rm \Delta{Y_{surf}}}$ & ${\rm t_{tip}~(Gyr)}$ & ${\rm M_{ZAHB}^{3.85} (M_\odot)}$ & ${\rm \log(L_{ZAHB}^{3.85}/L_\odot)}$\\
\hline                    
$10^{-5}$         & 0.245  & 5.35   & 2.477  & 3.183   & 0.5103 & 0.013 & 5.61  & $--$  & $--$  \\
$10^{-4}$         & 0.245  & 5.25   & 2.352  & 3.271   & 0.4972 & 0.016 & 5.57  & 0.801 & 1.757 \\
$3\times10^{-4}$  & 0.245  & 5.31   & 2.217  & 3.315   & 0.4924 & 0.018 & 5.69  & 0.707 & 1.710 \\
$6\times10^{-4}$  & 0.246  & 5.41   & 2.132  & 3.342   & 0.4895 & 0.019 & 5.84  & 0.665 & 1.685 \\
$10^{-3}$         & 0.246  & 5.58   & 2.024  & 3.361   & 0.4876 & 0.020 & 6.07  & 0.640 & 1.664 \\
$2\times10^{-3}$  & 0.248  & 5.96   & 1.915  & 3.387   & 0.4853 & 0.020 & 6.55  & 0.609 & 1.631 \\
$4\times10^{-3}$  & 0.251  & 6.72   & 1.757  & 3.412   & 0.4829 & 0.021 & 7.48  & 0.585 & 1.593 \\
$8\times10^{-3}$  & 0.256  & 8.09   & 1.608  & 3.433   & 0.4798 & 0.023 & 9.10  & 0.564 & 1.543 \\
$10^{-2}$         & 0.259  & 8.75   & 1.555  & 3.437   & 0.4785 & 0.022 & 9.84  & 0.558 & 1.521 \\
0.0198            & 0.2734 & 11.29  & 1.391  & 3.446   & 0.4727 & 0.023 & 12.61 & 0.538 & 1.461 \\
0.03              & 0.288  & 12.54  & 1.318  & 3.447   & 0.4675 & 0.023 & 13.94 & 0.526 & 1.431 \\
0.04              & 0.303  & 13.07  & 1.297  & 3.441   & 0.4621 & 0.021 & 14.49 & 0.516 & 1.419 \\
0.05              & 0.316  & 13.18  & 1.279  & 3.436   & 0.4574 & 0.021 & 14.55 & 0.508 & 1.417 \\

\hline                  
\end{tabular}
\end{table*}

\begin{figure}
 \centering
 \includegraphics[scale=0.45]{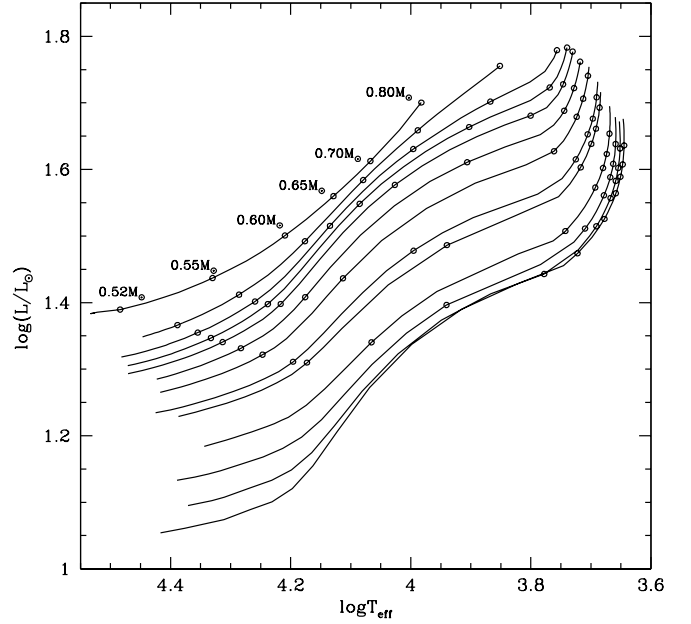}
 \caption{HR diagram of ZAHBs for the complete set of scaled solar metallicities (Z increases from top to bottom). Filled circles 
correspond to selected models with ${\rm M/M_\odot}$: 0.52, 0.55, 0.60, 0.65, 0.70 and 0.80.}
\label{fig:zahbzz}
\end{figure}

As already mentioned, Fig.~\ref{fig:hb} displays the evolutionary tracks of selected  models 
along both core and shell He-burning stages, for the new chemical compositions presented in this paper. 
They originate from  
a $\sim$13 Gyr old RGB progenitor, with an initial mass ${\rm M=0.8M_\odot}$ for ${\rm Z=10^{-5}}$, 
and ${\rm 1M_\odot}$ for Z=0.05, respectively. 
Super-metal-rich 
HB stars have a cooler ZAHB location compared to the extremely metal-poor counterpart with the same total mass, due 
to the higher envelope opacity and smaller He core. Metal-rich HB models need to 
experience a huge mass 
loss along the RGB to be able to reach ZAHB effective temperatures hotter than $\sim10,000$~K:
for instance for Z=0.05, the stellar model whose ZAHB location is at ${\rm T_{eff}=10,000}$~K has a total mass equal to ${\rm \approx0.50M_\odot}$
corresponding to an amount of mass lost by the RGB progenitor equal to ${\rm \Delta{M}\approx0.5M_\odot}$, i.e. about 50\%  of the initial mass.
On the other hand, all the very metal-poor models, regardless of the amount of mass lost during the RGB, have a ZAHB location 
hotter than the blue boundary of the RR Lyrae instability strip. 
This implies that both extremely metal-poor and  
super-metal-rich stellar populations have a very small -- if any -- probability 
to produce significant populations of RR Lyrae variable stars 
(see also Cassisi et al.~1997, and Bono et al.~1997). 

Figure~\ref{fig:zahbzz} shows the zero age HBs (ZAHBs) for all BaSTI scaled solar compositions. 
They obviously become progressively fainter and cooler with increasing metallicity, due -- as already mentioned -- 
to larger envelope opacities, but also -- and mainly -- to the smaller He 
core at He ignition. 
It is worth noticing the behaviour of the Z=0.05 ZAHB, whose brightness 
at ${\rm T_{eff}<10,000}$~K coincides with that of the Z=0.04 counterpart. This can be explained as follows.
The ZAHB brightness depends on both the mass of the He core 
and the shell H-burning efficiency, this latter controlled by both mass and chemical composition of the 
envelope. At the hot end of the 
ZAHB the envelope mass is tiny, the H-burning efficiency is vanishing, and therefore the ZAHB luminosity is 
only a function of the He core mass. 
Given that ${\rm M_{cHe}}$ is smaller at Z=0.05 compared to Z=0.04, the ZAHB models are 
correspondingly fainter. 
When increasing the envelope mass (that is, moving towards the cool side of the ZAHB) 
shell burning provides a major contribution to the energy budget. Given that Z=0.05 models have a 
larger envelope He (see data in Tab.~\ref{tab:1mevo}) and CNO abundances, the shell burning is more efficient 
and compensates for the smaller He core mass. As a consequence, 
the ZAHB for Z=0.04 and Y=0.303 overlaps the ZAHB for Z=0.05 and Y=0.316 below a given ${\rm T_{eff}}$.

\begin{figure}
 \centering
 \includegraphics[scale=0.45]{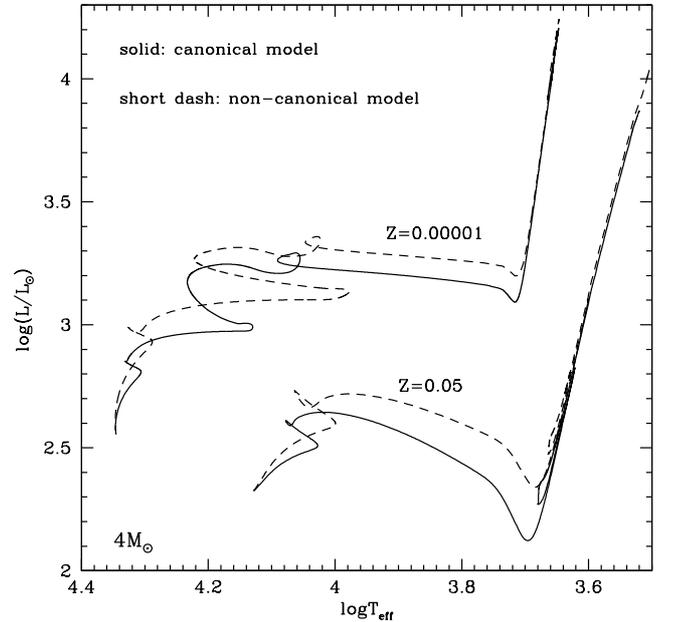}
 \caption{Evolutionary tracks for ${\rm 4.0M_\odot}$ stellar models computed with (non canonical) 
and without (canonical) MS convective core overshooting, for both ${\rm Z=10^{-5}}$ and 0.05.}
\label{fig:4ssov}
\end{figure}

As for the intermediate-mass models, some relevant evolutionary and 
structural properties of representative ${\rm 4M_\odot}$ 
models computed with and without MS convective core overshooting,
 for all scaled solar metallicities of the BaSTI archive, are listed in 
Table~\ref{tab:4mevo}. A comparison between canonical (no overshooting) 
and non-canonical (with overshooting) evolutionary 
tracks of a ${\rm 4M_\odot}$ stellar model, for ${\rm Z=0.00001}$ and 0.05, is shown in Fig.~\ref{fig:4ssov}.

It is important to notice that the strong dependence of the radiative opacity 
on metals affect the properties of the intermediate-mass models in the same way as for the low-mass ones. 
Tracks shown in Fig.~\ref{fig:trkiso} 
reveal that at ${\rm Z=10^{-5}}$ the transition from lower (H-burning occurring mainly via the p-p 
chain) to upper  (dominated by the CNO cycle) MS stars happens at ${\rm \sim 2.3M_\odot}$, as 
testified by the appearance of the overall contraction phase at central H exhaustion. 
For Z=0.05 the transition occurs at  ${\rm \sim1.1M_\odot}$. 
This huge difference is due to the fact that by decreasing the metallicity, the abundance of CNO elements also decreases, 
and one needs much higher central temperatures (hence larger masses) for the CNO cycle 
energy generation to dominate over the p-p chain contribution.  

\begin{figure}
 \centering
 \includegraphics[scale=0.45]{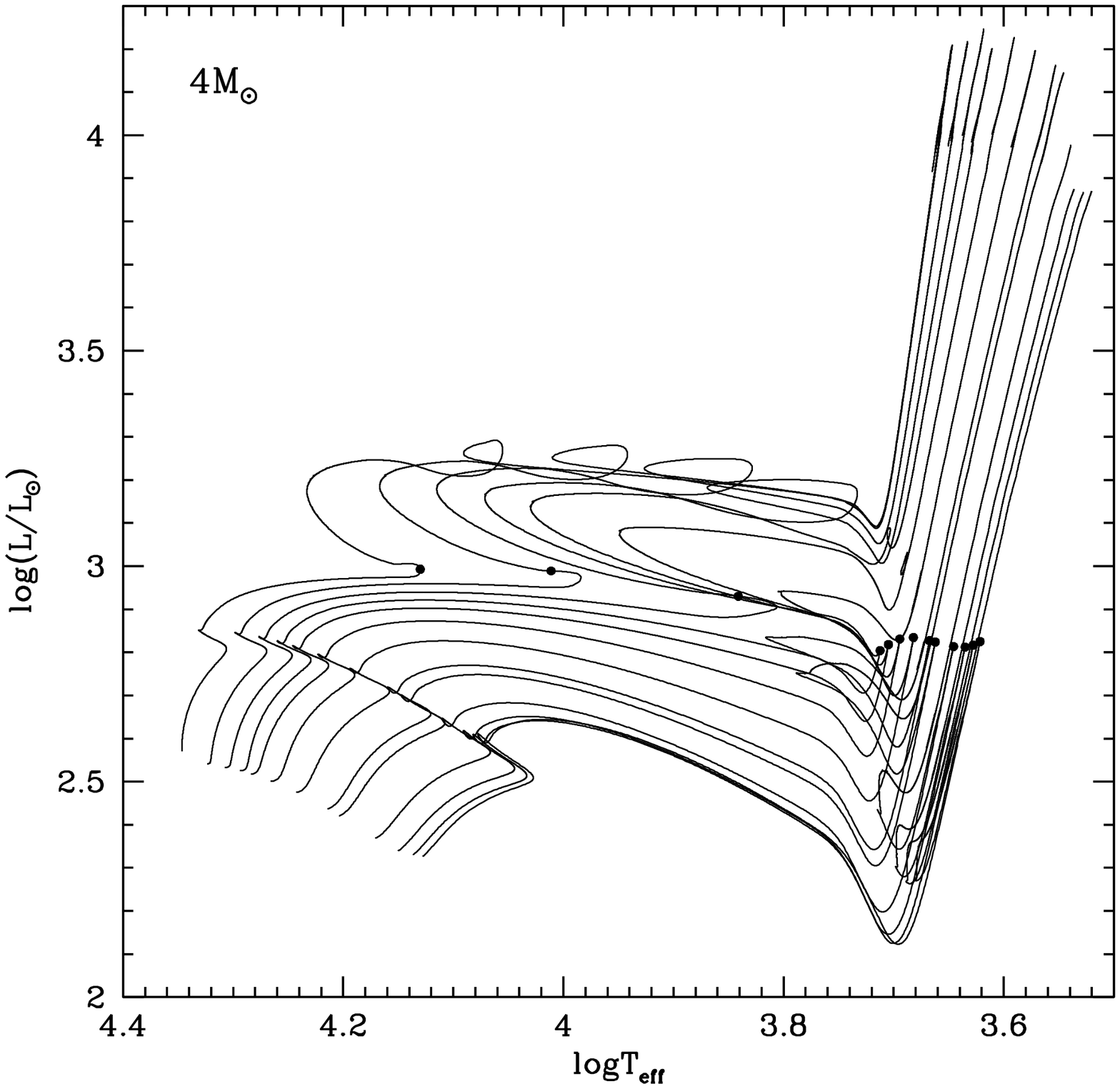}
 \caption{Evolutionary tracks for ${\rm 4.0M_\odot}$ models (no overshooting) and all scaled solar chemical compositions 
of the BaSTI database. Filled circles mark central He-burning ignition.}
\label{fig:4mzz}
\end{figure}

The generally higher central temperatures of the metal poor models also imply that for intermediate-mass models the 
thermal conditions for He-burning ignition are reached earlier than their 
metal-rich counterparts. Therefore, at very low Z, intermediate-mass models miss the RGB stage as shown in 
Fig.~\ref{fig:4mzz}. 

\begin{table*}
\caption{Selected evolutionary properties for ${\rm 4M_\odot}$ models computed 
with and without MS convective core overshooting, for all scaled-solar BaSTI  
metallicities: ${\rm M_{cc}}$ is the convective core mass at the beginning of central 
H-burning; ${\rm t_H}$  corresponds
to the age at the central H exhaustion; ${\rm M_{cHe}(M_\odot)}$ and ${\rm logT_{eff}}$ are the helium core mass and the 
logarithm of the effective temperature 
at the ignition of central He-burning, respectively; ${\rm t_{He}}$ is the core helium burning lifetime;  
${\rm \Delta{Y_{surf}}}$ is the amount of extra helium dredged to the surface
during the ${\rm 2^{nd}}$ dredge-up; ${\rm M_{cCO}(M_\odot)}$ is the C-O core mass at the ${\rm 1^{st}}$ thermal pulse.}             
\label{tab:4mevo}      
\centering          
\begin{tabular}{c c c c c c c c c}     
\hline\hline       
                      
Z  & Y &${\rm M_{cc}(M_\odot)}$ & ${\rm t_{H}(Myr)}$  & ${\rm M_{cHe}(M_\odot)}$  & ${\rm logT_{eff}}$ & ${\rm t_{He}(Myr)}$ & ${\rm \Delta{Y_{surf}}}$ & ${\rm M_{cCO}(M_\odot)}$\\
\hline                    
\multicolumn{9}{c}{no MS core overshooting}\\
\hline
$10^{-5}$         & 0.245  & 0.9769 & 101.41  &  0.5256  & 4.154 & 24.42 & 0.062 & 0.8225 \\
$10^{-4}$         & 0.245  & 1.0255 & 106.64  &  0.5282  & 4.038 & 24.08 & 0.059 & 0.8248 \\
$3\times10^{-4}$  & 0.245  & 1.0356 & 109.21  &  0.5282  & 3.937 & 24.80 & 0.056 & 0.8234 \\
$6\times10^{-4}$  & 0.246  & 1.0376 & 114.10  &  0.5232  & 3.817 & 26.31 & 0.050 & 0.8209 \\
$10^{-3}$         & 0.246  & 1.0405 & 112.62  &  0.5193  & 3.731 & 28.26 & 0.048 & 0.8190 \\
$2\times10^{-3}$  & 0.248  & 1.0505 & 116.06  &  0.5123  & 3.699 & 31.29 & 0.036 & 0.8131 \\
$4\times10^{-3}$  & 0.251  & 1.0117 & 119.71  &  0.4993  & 3.686 & 34.65 & 0.022 & 0.7978 \\
$8\times10^{-3}$  & 0.256  & 0.9751 & 126.35  &  0.4891  & 3.672 & 37.54 & 0.008 & 0.7846 \\
$10^{-2}$         & 0.259  & 0.9751 & 129.61  &  0.4854  & 3.667 & 38.75 & 0.005 & 0.7780 \\
0.0198            & 0.2734 & 0.9221 & 140.34  &  0.4812  & 3.650 & 45.87 & no $2^{nd}$ DU & 0.7143 \\
0.03              & 0.288  & 0.8850 & 144.58  &  0.4851  & 3.640 & 49.11 & no $2^{nd}$ DU & 0.6840 \\
0.04              & 0.303  & 0.8459 & 143.61  &  0.4930  & 3.633 & 49.12 & no $2^{nd}$ DU & 0.6823 \\
0.05              & 0.316  & 0.8374 & 142.80  &  0.5052  & 3.627 & 46.52 & no $2^{nd}$ DU & 0.6971 \\
\hline
\multicolumn{9}{c}{MS core overshooting}\\
\hline

$10^{-5}$         & 0.245  & 0.9808 & 124.25  &  0.6208  & 4.053 & 15.42 & 0.085 & 0.8477 \\
$10^{-4}$         & 0.245  & 1.0367 & 129.68  &  0.6329  & 3.843 & 15.54 & 0.087 & 0.8557 \\
$3\times10^{-4}$  & 0.245  & 1.0391 & 132.13  &  0.6315  & 3.712 & 16.21 & 0.085 & 0.8516 \\
$6\times10^{-4}$  & 0.246  & 1.0406 & 133.81  &  0.6272  & 3.696 & 16.80 & 0.082 & 0.8510 \\
$10^{-3}$         & 0.246  & 1.0427 & 135.80  &  0.6219  & 3.689 & 17.49 & 0.079 & 0.8465 \\
$2\times10^{-3}$  & 0.248  & 1.0265 & 142.76  &  0.6114  & 3.678 & 18.02 & 0.069 & 0.8365 \\
$4\times10^{-3}$  & 0.251  & 1.0056 & 145.64  &  0.5959  & 3.665 & 19.06 & 0.059 & 0.8290 \\
$8\times10^{-3}$  & 0.256  & 0.9672 & 152.68  &  0.5814  & 3.651 & 21.18 & 0.054 & 0.8108 \\
$10^{-2}$         & 0.259  & 0.9566 & 158.80  &  0.5774  & 3.645 & 23.51 & 0.052 & 0.8014 \\
0.0198            & 0.2734 & 0.9082 & 171.17  &  0.5673  & 3.629 & 28.93 & 0.041 & 0.7787 \\
0.03              & 0.288  & 0.8781 & 175.48  &  0.5703  & 3.618 & 31.25 & 0.037 & 0.7696 \\
0.04              & 0.303  & 0.8583 & 174.45  &  0.5796  & 3.611 & 29.64 & 0.035 & 0.7659 \\
0.05              & 0.316  & 0.8487 & 170.05  &  0.5903  & 3.605 & 28.85 & 0.035 & 0.7655 \\

\hline                  
\end{tabular}
\end{table*}

\begin{figure}
 \centering
 \includegraphics[scale=0.44]{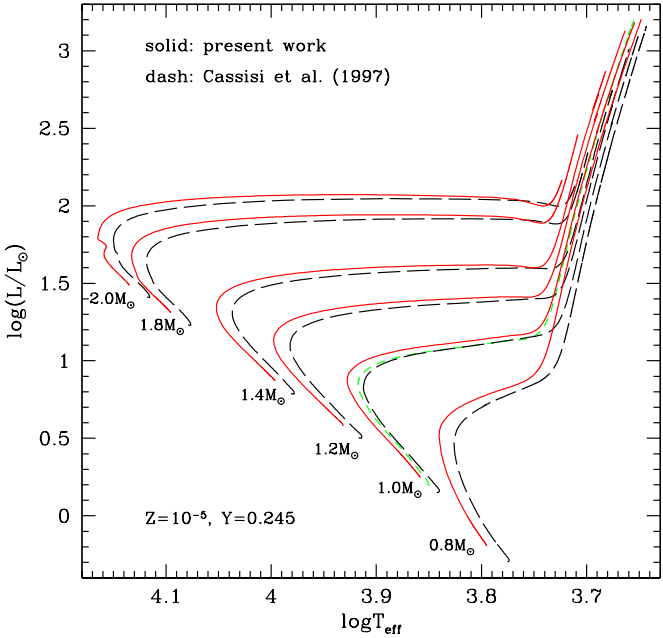}
 \caption{HR diagram of selected H-burning models compared to Cassisi et al.~(1997) results. 
The short dashed line corresponds to a ${\rm 1M_\odot}$ model 
computed with our updated physical inputs but the same initial He content adopted by Cassisi et al.~(1997).}
 \label{fig:trkc97}
\end{figure}

\section{Comparison with observations and literature models}

Unfortunately, there are not many observations that sample the two extreme metallicities discussed in this paper, 
and the available ones are not accurate enough to enable stringent tests of the models.

\begin{figure}
 \centering
 \includegraphics[scale=0.44]{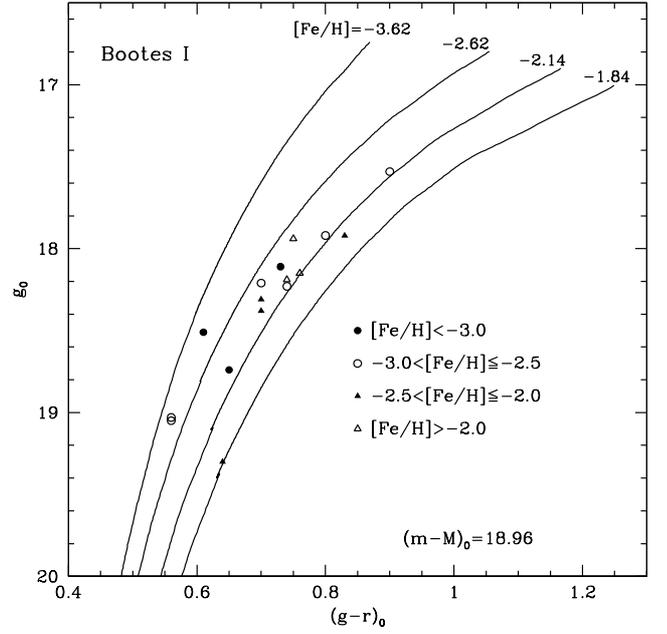}
 \caption{Colour Magnitude Diagram of a sample of metal poor RGB stars belonging to the Bo\"otes~I 
dwarf spheroidal galaxy, that have spectroscopic [Fe/H] determinations (Norris et al.~2008). 
Theoretical isochrones for an age of 12.5~Gyr and 
the labelled [Fe/H] values (for an $\alpha$-enhanced metal mixture) are also displayed 
(see text for details).}
 \label{fig:bootes1}
\end{figure}

We show in Fig.~\ref{fig:bootes1} a comparison of the CMD of selected RGB stars in the 
metal poor Bo\"otes~I dwarf spheroidal galaxy (Norris et al.~2008), with 12.5~Gyr $\alpha$-enhanced 
theoretical isochrones and various [Fe/H] values (the precise value of the age is not critical). 
The observational points have [Fe/H] determinations 
in the range between [Fe/H]$\sim -$1.5 and [Fe/H]$\sim -$3.5, with standard deviations 
on the individual measurements between $\sim$0.19 and $\sim$0.35 dex.
A distance modulus ${\rm (m-M)_0}=18.96$ from Kuehn et al.~(2008) has been applied to the isochrones. 
Observational data were dereddened by Norris et al.~(2008) using E(B-V)=0.02.
Taking into account the substantial error on the [Fe/H] estimates, there is a general broad agreement 
between models and observations.

\begin{figure}
 \centering
 \includegraphics[scale=0.44]{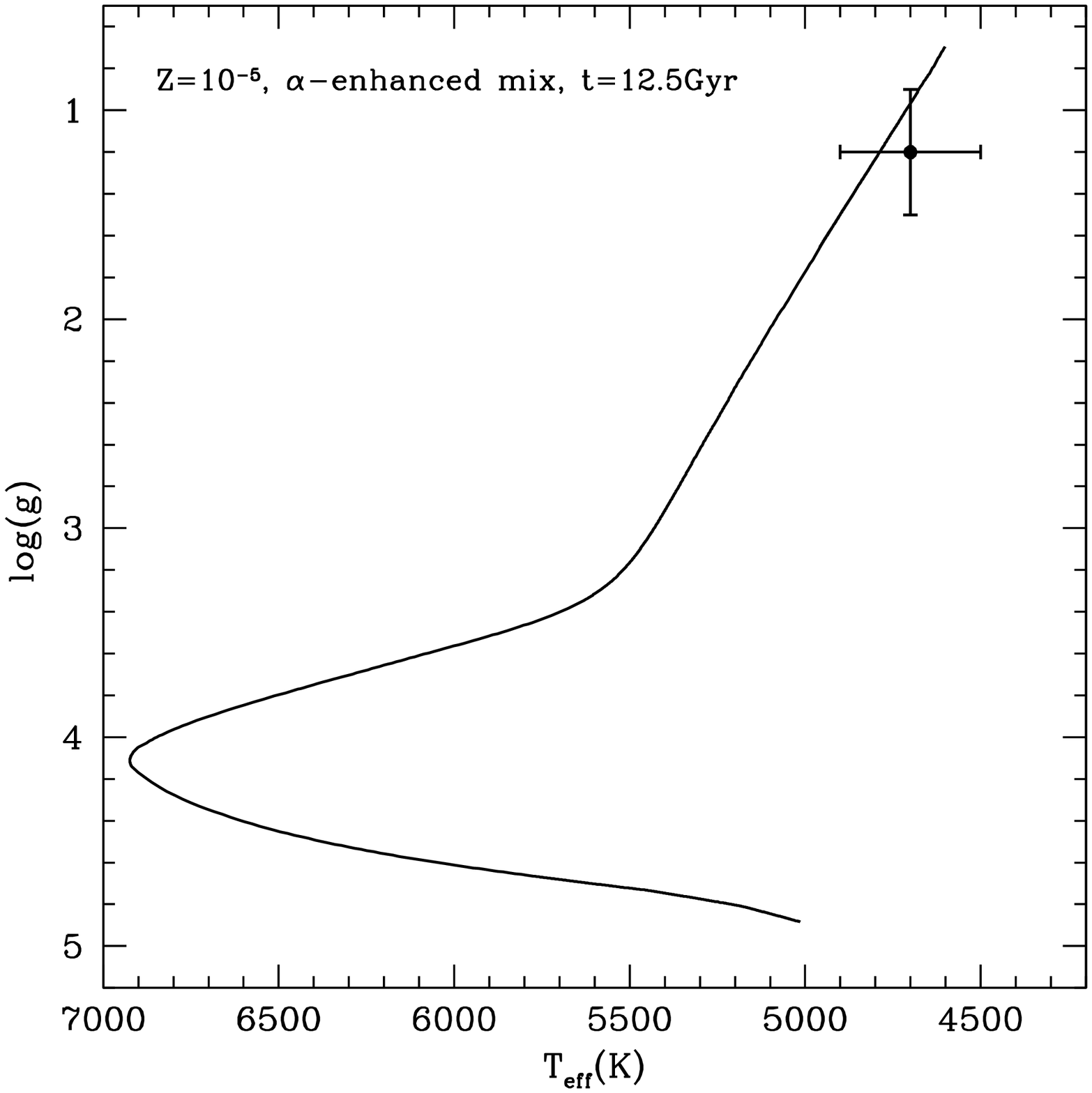}
 \caption{Comparison of a 12.5~Gyr, [Fe/H]$-$3.62 $\alpha$-enhanced isochrone in the log(g)${\rm T_{eff}}$ diagram, with 
the extremely metal poor ([Fe/H]=$-$3.7$\pm$0.1) star Boo-1137, in the Bo\"otes I dwarf spheroidal galaxy 
(Norris et al. 2010).}
 \label{fig:bootes2}
\end{figure}

Figure~\ref{fig:bootes2} compares the log(g)-${\rm T_{eff}}$ diagram of a 12.5~Gyr, [Fe/H]$-$3.62 
(${\rm Z=10^{-5}}$) $\alpha$-enhanced isochrone, 
with the extremely metal poor star Boo-1137, in the same dwarf spheroidal galaxy. This star has a spectroscopic determination of 
[Fe/H]=$-$3.7$\pm$0.1 (and $\alpha$-element enhancement -- Norris et al.~2010).
Error bars quoted by the authors are also displayed.
Within the error bars, the extremely metal poor models discussed in this paper are able to match  
the observations. 

To the best of our knowledge, stellar models for these extreme metallicities have been calculated by few authors.
Models for ${\rm Z=10^{-5}}$, covering both low- and intermediate-mass stars have been 
computed by Cassisi et al.~(1997), and can be also found as part of the Yale-Yonsei (YY) model database (see, e.g., 
Demarque et al.~2004). Figure~\ref{fig:mchez15}
shows the comparison between our models (without convective overshooting, as in Cassisi et al.~1997 calculations) 
and Cassisi et al.~(1997) results, 
concerning both the He core mass and surface luminosity at the RGB tip, 
for stellar masses around the RGB transition, while Fig.~\ref{fig:trkc97} compares the HR diagrams of selected H-burning models.  

The differences of the evolutionary tracks are due to the the completely different physics inputs adopted, but 
we need also to consider that Cassisi et al.~(1997) computations 
assume an initial Y=0.23, while we now employ an updated and higher estimate of the primordial 
He abundance (Y=0.245). To isolate this latter effect, we have performed an additional calculation for a ${\rm 1M_\odot}$ model, 
by employing BaSTI physics inputs, but Y=0.23, as in Cassisi et al.~(1997).
This evolutionary track is shown by Fig.~\ref{fig:trkc97} 
to be very close to the old computations along the MS, while matching along the RGB the new BaSTI results. 
This implies that along the RGB the difference with Cassisi et al.~(1997) models is due to 
the different low temperature radiative opacities, while it is the different initial He abundance that mainly causes the   
differences along the MS and subgiant branch. 

\begin{figure}
 \centering
 \includegraphics[scale=0.44]{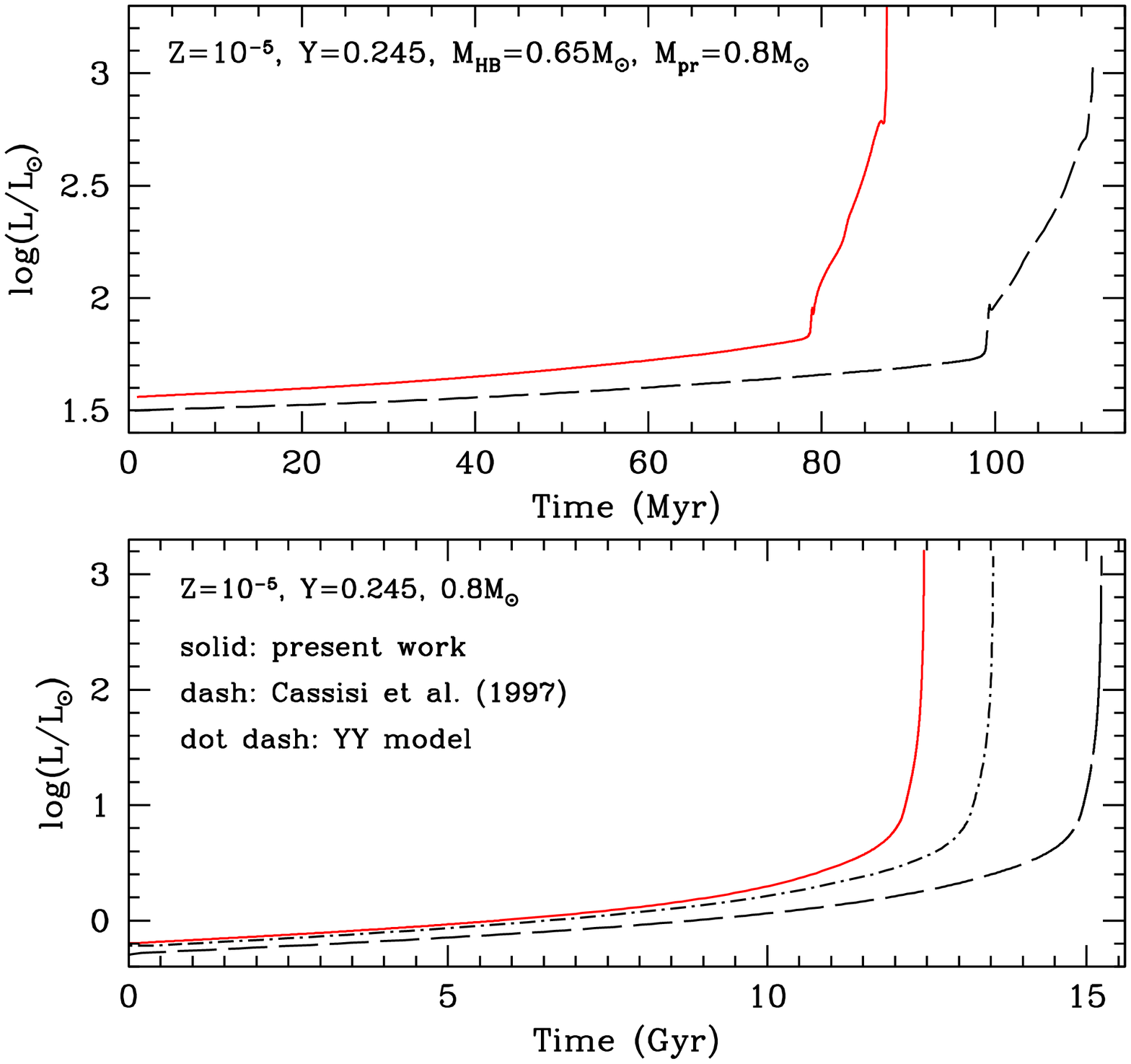}
 \caption{\textit{Upper panel}: the behaviour of the surface luminosity as a function of time for two  
HB models with the same mass (see labels). 
The solid line corresponds to our computations, while the dashed line displays results 
by Cassisi et al.~(1997). 
\textit{Lower panel}: as the upper panel, but for 0.8${\rm M_{\odot}}$ calculations from the zero age MS to the RGB tip, including also a YY model.}
 \label{fig:timec97}
\end{figure}

\begin{figure}
 \centering
 \includegraphics[scale=0.44]{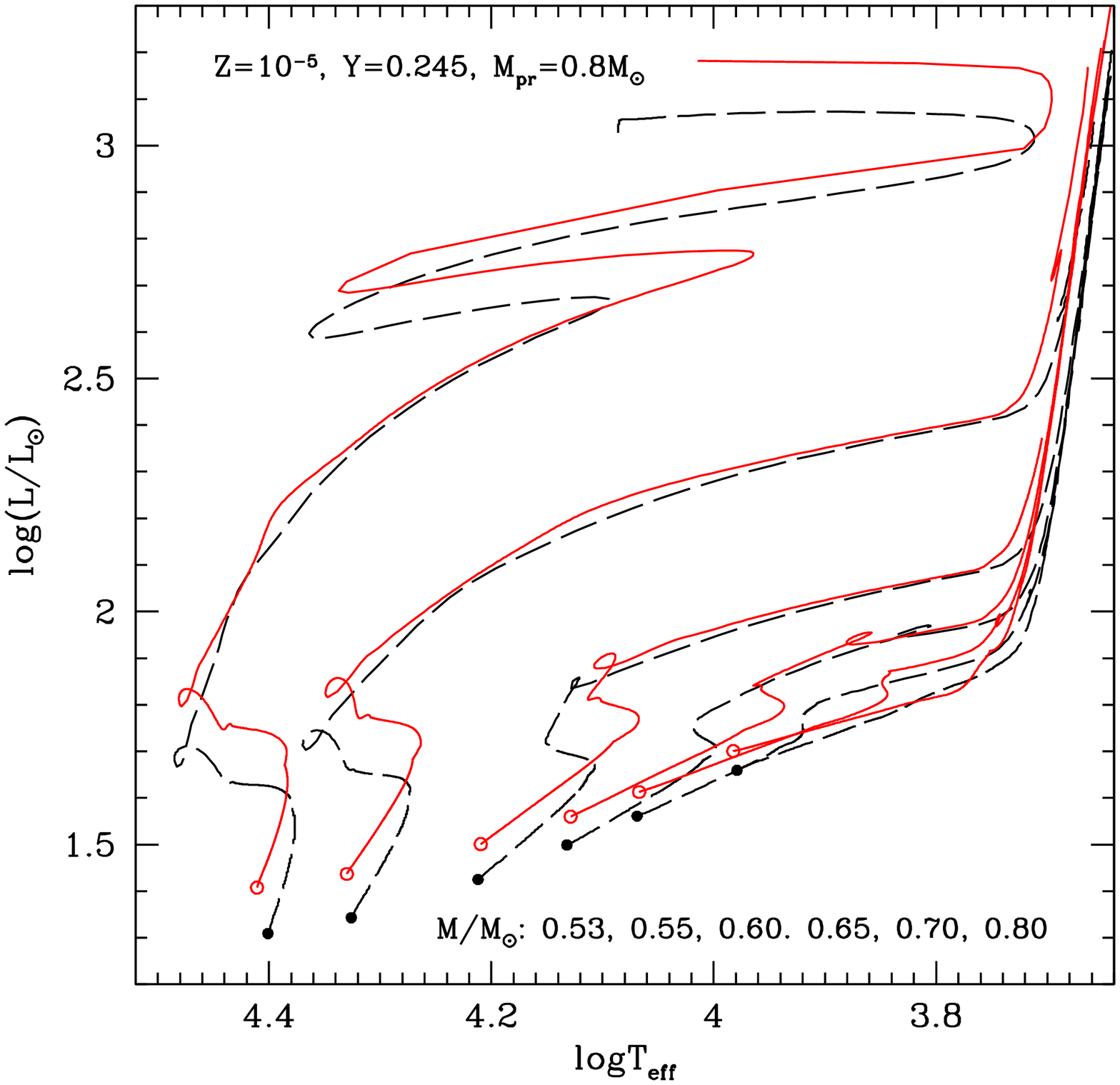}
 \caption{As Fig.~\ref{fig:timec97} but for the  HR diagram of selected HB models with the same RGB progenitor mass (${\rm M_{pr}}$). }
 \label{fig:hbc97}
\end{figure}

As for the evolutionary lifetimes, the lower panel of Fig.~\ref{fig:timec97} compares the 0.8${\rm M_{\odot}}$ 
luminosity-age relation with the Cassisi et al.~(1997) counterpart. 
The MS lifetime is shorter by about 2.5~Gyr in  
our computations, a consequence of both the larger initial He content and different physics inputs (a fundamental 
role is played by the different equation of state). The upper panel of the same figure compares 
the trend of the  
luminosity with time of a HB model originated from 
a ${\rm 0.8M_\odot}$ RGB progenitor, with the Cassisi et al.~(1997) counterpart. 
Our model has a core He-burning 
lifetime shorter by about 20\% with respect to the older calculations. This results is a consequence of different 
physics inputs, in particular the use of an updated nuclear cross section for 
the ${\rm ^{12}C(\alpha,\gamma)^{16}O}$ reaction (Kunz et al.~2002).

HR diagrams of selected HB models are compared in Fig.~\ref{fig:hbc97}. There is 
a relatively small luminosity difference that is 
mainly a consequence of the different envelope He abundance, for the
RGB progenitors have a similar He core mass at the RGB tip (see data in Fig.~\ref{fig:mchez15}).

\begin{figure}
 \centering
 \includegraphics[scale=0.44]{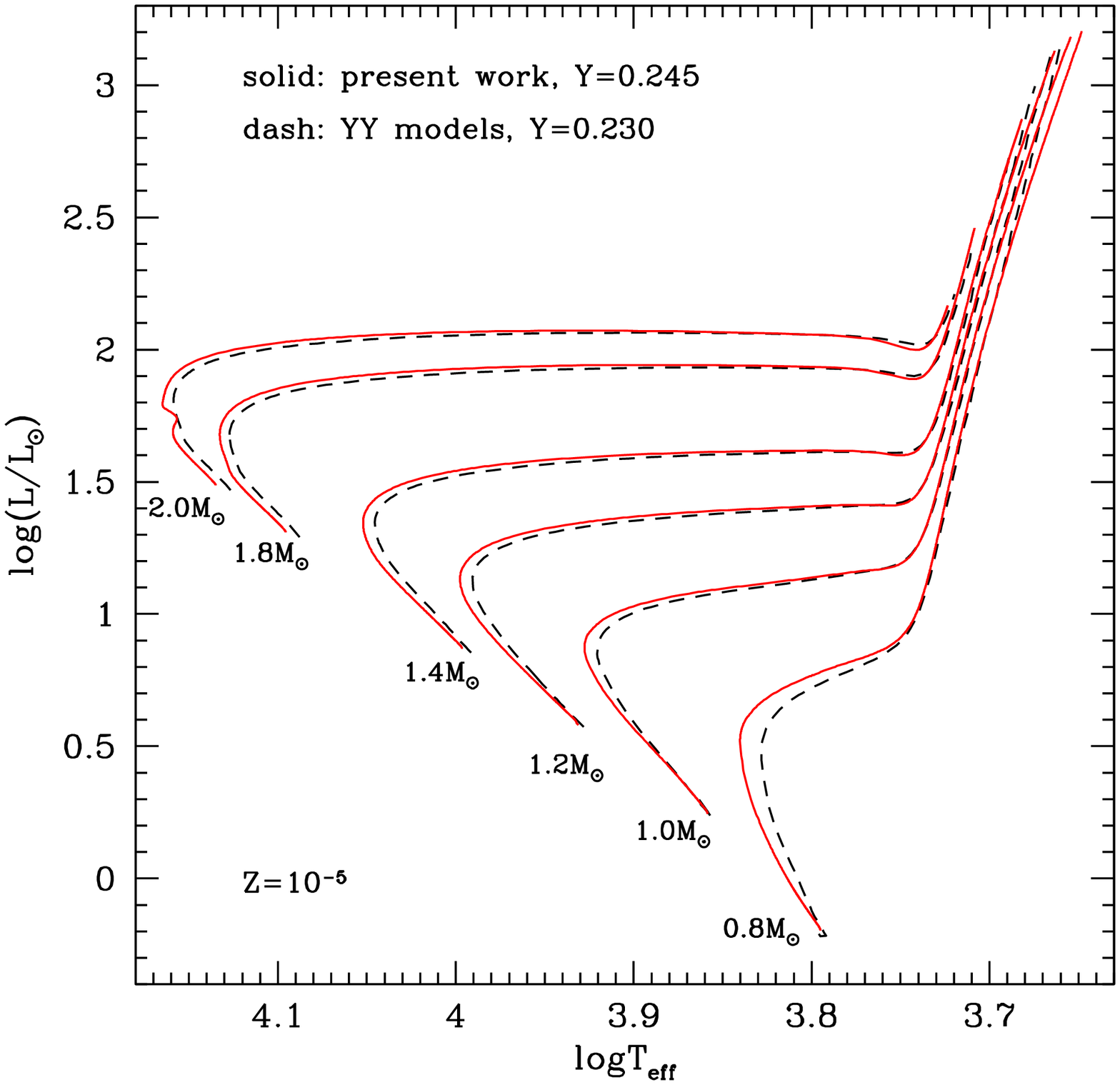}
 \caption{As Fig.~\ref{fig:trkc97}, but for a comparison with YY models.}
 \label{fig:YY}
\end{figure}

Figure~\ref{fig:YY} displays a comparison similar to Fig.~\ref{fig:trkc97}, but with YY models. 
Here we consider BaSTI models with convective core overshooting, because it is included 
(in a similar way as BaSTI) in the YY calculations.
This latter set adopts physics inputs very similar to our calculations, the main differences being 
the inclusion of He diffusion (that affects the lower mass models) and a lower initial He abundance, Y=0.23.
For masses above $\sim$1.2${\rm M_{\odot}}$ the differences in the HR diagrams of the two sets of 
tracks are due essentially to the different initial Y, 
while for the lower masses the effect of He diffusion amplifies the differences around the turn off (TO).
The RGBs of the two sets of calculations are almost identical, for they are very weakly affected by the initial Y and 
efficiency of atomic diffusion.
Evolutionary lifetimes during the core and shell H-burning stages for the 0.8${\rm M_{\odot}}$ YY model are 
displayed in the lower panel of Fig.~\ref{fig:timec97}; they turn out to be 
longer by $\sim$1~Gyr for the MS, compared to BaSTI models, in spite of the efficient atomic diffusion 
that tends to shorten the MS lifetime compared to the no-diffusion case. 
The main reason for this difference is very likely the higher initial He abundance in BaSTI models that, as mentioned already, 
decreases the MS lifetime. 

\begin{figure}
 \centering
 \includegraphics[scale=0.44]{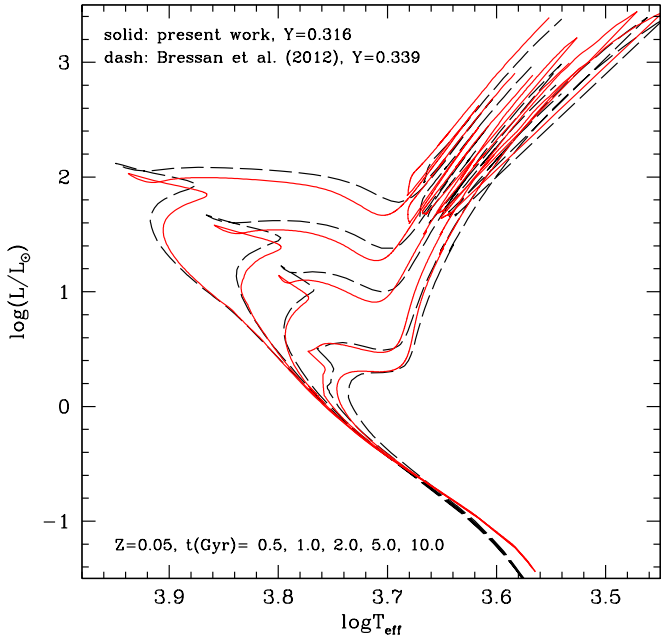}
 \caption{Selected isochrones from BaSTI and the PARSEC databases for Z=0.05. 
The BaSTI isochrones include convective core overshooting during the central H-burning stage.}
 \label{fig:isobastipd}
\end{figure}

As for the Z=0.05 models, 
the only possibility we have for comparisons with independent calculations at exactly the same Z, is 
to consider the models by Bressan et al.~(2012, PARSEC models). The PARSEC 
library includes a grid point at Z=0.05, although with a larger initial He content. 
The physics inputs of PARSEC calculations are similar to those 
adopted for the BaSTI database, the main differences being the low-T and electron conduction 
opacities, and some nuclear reaction rates, plus their inclusion of atomic diffusion. 
Their adopted scaled solar heavy element mixture starts from 
Grevesse \& Sauval~(1998), but supplemented for a subset of 
elements by Caffau et al.~(2011, and references therein) results. 
In particular, the very abundant CNO elements and Fe are amongst the metals with 
Caffau et al. (2011) abundances, and the resulting metal mixture is appreciably different from 
our calculations.
Figure~\ref{fig:isobastipd} displays a comparison of selected isochrones from BaSTI and PARSEC, from the MS to the 
asymptotic giant branch phase.  
We display the BaSTI results including core overshooting, because of its inclusion in PARSEC calculations. 

The two sets of isochrones display differences along the various branches, that are not too large but still 
noticeable. For isochrones populated by stars with well developed convective cores along the MS 
(ages below 5~Gyr), PARSEC calculations 
display typically brighter and hotter TOs, while the reverse is true for the older ages. The RGB and asymptotic giant branch 
${\rm T_{eff}}$ of our models is typically larger (the reverse is true for the lower MS), 
and the central He burning luminosity is generally fainter.
It is difficult to disentangle the various causes of this differences, that are undoubtedly due to the different initial He 
(that however should not affect appreciably the RGB ${\rm T_{eff}}$, and would in any case exacerbate the differences along this phase), 
the efficiency of atomic diffusion (for the older isochrone TO region only), a slightly different extension of the overshooting region 
(larger in the PARSEC models) in stars with well developed MS convective cores and the different initial metal mixture.  
 
\section{Summary}

We have presented an extension of the BaSTI stellar evolution database to extremely metal poor (${\rm Z=10^{-5}}$) and 
super-metal-rich (Z=0.05) metallicities, with both scaled-solar and $\alpha$-enhanced ([$\alpha$/Fe]=0.4) 
heavy element distributions. With these new additions, the BaSTI database can be used to model stellar populations 
ranging from the most metal poor components of faint dwarf galaxies in the Local Group, to the metal rich 
populations of the Galactic bulge.

We have discussed the variations of several fundamental predictions of stellar evolution over the large 
metallicity range spanned by the full BaSTI models, and compared the new calculations with 
literature models at ${\rm Z=10^{-5}}$ and Z=0.05. 
The comparison discloses a good agreement with YY calculations from the MS to the tip of the RGB 
-- YY models employ very similar input physics -- 
at ${\rm Z=10^{-5}}$. The existing small differences in the HR diagram are easily explained by the different initial He abundance, and the 
inclusion of He diffusion in YY calculations.
As for the comparison with Cassisi et al.~(1997) ${\rm Z=10^{-5}}$ models, the ${\rm T_{eff}}$ difference 
along the RGB is due to 
the different low temperature radiative opacities, while it is the different initial He abundance that causes the 
differences in ${\rm T_{eff}}$ and bolometric luminosity along the MS and subgiant branch. 
Cassisi et al.~(1997) calculations include also HB models, that tend to be slightly fainter, mainly due to the lower initial Y.

The only existing modern calculations at Z=0.05 are from the PARSEC database, and we compared isochrones for 
selected ages between 0.5 and 10.0~Gyr. Several differences appear between the two sets, whose causes are difficult 
to disentangle. They are due to the different initial He, 
the inclusion of atomic diffusion in the PARSEC models (for the older isochrone TO region only), 
a slightly different extension of the overshooting region 
in stars with well developed MS convective cores and the different initial metal mixture.  

\begin{acknowledgements}

We warmly thank our referee for her/his pertinent comments that have improved the readability of an
early version of the manuscript. SC is grateful for financial support 
from PRIN-INAF 2011 "Multiple Populations in Globular Clusters: their 
role in the Galaxy assembly" (PI: E. Carretta), and from PRIN MIUR 2010-2011, 
project \lq{The Chemical and Dynamical Evolution of the Milky Way and Local Group Galaxies}\rq, prot. 2010LY5N2T (PI: F. Matteucci).
This research has made use of NASA's Astrophysics Data System Abstract
Service and the SIMBAD database operated at CDS, Strasbourg, France.

\end{acknowledgements}



\end{document}